\documentclass[a4paper,12pt]{article}
\usepackage{amssymb,amsmath}
\usepackage{bm}

\usepackage{cite}

\def\nd{\end{document}}
\def\rg{\rangle} \def\lg{\langle}
\def\ha{{\textstyle{1\over2}}}
\def\bbz{\mathbb{Z}}
\def\bbr{\mathbb{R}}

\def\hC{\hat C}

\def\vf{\varphi}
\def\D{\Delta} 
\def\L{\Lambda} \def\l{\lambda} \def\z{\zeta}
 \def\b{\beta} \def\s{{\sigma}}
\def\Ga{\Gamma} \def\g{\gamma}

\def\beq{\begin{equation}}
\def\eqn#1{\beq\label{#1}}
\def\eeq{\end{equation}}
\def\bea{\begin{eqnarray}}
\def\eea{\end{eqnarray}}
 
\def\eqnn#1{\bea\label{#1}}

\def\nn{\nonumber}

\def\del#1{ \partial_{#1} }
\def\s#1{{\mathfrak s}_{#1}}
\def\Del#1{ \frac{\partial}{\partial #1} }
\def\Cas{{\cal C}} \def\cg{{\cal G}}
\def\G{\mathsf{G}}
\def\np{\newpage}
\def\lra{\leftrightarrow}
\def\ra{\rightarrow}

\def\ca{{\cal A}}  \def\cc{{\cal C}}
  
\def\cg{{\cal G}} \def\ch{{\cal H}} 
 \def\ck{{\cal K}} \def\cl{{\cal L}}
\def\cm{{\cal M}} \def\cn{{\cal N}} 
\def\cp{{\cal P}} \def\cq{{\cal Q}}

\def\tcn{\widetilde{{\cal N}}}
\def\tN{\widetilde{N}}
\def\tn{\widetilde{n}}
\def\tu{\tilde{u}}
\def\tI{\tilde{I}} \def\tL{{\tilde{\L}}}

\def\hu{\hat{u}}
\def\tD{{\tilde{\D}}}

\begin{document}
\begin{flushright}
\end{flushright}
\begin{center}

 \textsf{\LARGE Intertwining Operator Realization of\\[2mm]
anti de Sitter  Holography}

\vspace{7mm}

{\large N.~Aizawa$^{a,}$\footnote{aizawa@mi.s.osakafu-u.ac.jp}, ~~~
V.K.~Dobrev$^{b,}$\footnote{dobrev@inrne.bas.bg}
 }

\vspace{3mm}

 \textit{$^a$ Department of Mathematics and Information Sciences,\\ Graduate School of Science,
Osaka Prefecture University, Nakamozu Campus,\\ Sakai, Osaka 599-8531 
Japan} \vspace{3mm}

 \textit{$^b$Institute of Nuclear Research and Nuclear Energy,\\
 Bulgarian Academy of Sciences \\
72 Tsarigradsko Chaussee, 1784 Sofia, Bulgaria}

\end{center}

\vspace{.8 cm}

\begin{abstract}
We give a group-theoretic interpretation of  relativistic holography
as equivalence between representations of the anti de Sitter algebra
describing  bulk fields  and  boundary fields. Our main result is
the explicit construction of the boundary-to-bulk operators  for
arbitrary integer spin  in the framework of representation theory.
Further we show that these operators and the bulk-to-boundary
operators are intertwining operators. In analogy to the de Sitter
case, we show that each bulk field has two boundary (shadow) fields
with conjugated conformal weights. These fields are related by
another intertwining operator given by a two-point function on the
boundary. \\

\end{abstract}

\vfill\newpage \pagestyle{plain} 


\setcounter{equation}{0}
\section{Introduction}

For the last fifteen years due to remarkable proposal of \cite{Malda}
the AdS/CFT correspondence is a dominant subject
in string theory and conformal field theory.
Actually the possible relation of field theory on anti de Sitter space
to conformal field theory on boundary Minkowski space-time was studied also before, cf.,
e.g., \cite{FlFr,AFFS,Fro,BrFr,NiSe,FeFr}.
The proposal of \cite{Malda} was further
elaborated in \cite{GKP} and \cite{Wita}.
After these initial papers there was an explosion of
related research which continues also currently, cf. e.g., \cite{FFZ}-\!\!\cite{HuLi}.

Let us remind that the AdS/CFT correspondence
has 2 ingredients \cite{Malda,GKP,Wita}:
1. the holography principle, which is very
old, and  means the reconstruction of some objects in the bulk (that
may be classical or quantum) from some objects on the boundary; 2.
the reconstruction of quantum objects, like 2-point functions on the
boundary, from appropriate actions on the bulk.

Our focus is on the first ingredient and we consider explicitly the simplest
case of the (3+1)-dimensional bulk with boundary 3D Minkowski space-time.
The reason for this is that until now the explicit presentation of
the holography principle was realized in the Euclidean case, relying on
Wick rotations of the final results, cf. e.g. \cite{Wita,Dobads}.

Yet it is desirable to  show the holography principle
by direct construction in Minkowski space-time.
This is what we do in the present paper    using representation theory only.
For this  we use a method that is used in the mathematical
literature for the construction of discrete series representations
of real semisimple Lie groups \cite{Hotta,Schmid}, and which method
was applied in the physics literature first in \cite{DMPPT}
exactly in the Euclidean AdS/CFT setting, though that term was not used then.\footnote{This method
was applied recently also to the case of non-relativistic holography \cite{AiDo}.}

The method  utilizes the fact that in the bulk the Casimir operators
are not fixed numerically. Thus, when a vector-field realization of
the anti de Sitter algebra ~$so(3,2)$~ is substituted in the bulk
Casimir it turns into a  differential  operator.  In contrast, the
boundary Casimir operators are fixed by the quantum numbers of the
fields under consideration. Then the bulk/boundary correspondence
forces an eigenvalue equation involving the Casimir differential
operator. That eigenvalue equation is used to find the two-point
Green function in the bulk which is then used to construct the
boundary-to-bulk integral operator. This operator maps a boundary
field   to a bulk field. This is our main result. We
stress that in our construction the bulk and boundary fields have
arbitrary integer spin. This is in sharp contrast to   preceding
results in the literature which considered spin zero and some very
low spin values  (in the Euclidean case
the intertwiners with arbitrary integer spin have been discussed in \cite{CDWW}).

What is also important in our approach is that we show that this
operator is an intertwining operator, namely, it intertwines the two
representations of the anti de Sitter algebra ~$so(3,2)$~ acting in the bulk and
on the boundary.

This also helps us to establish that each bulk
field has actually two bulk-to-boundary limits. The two boundary
fields have conjugated conformal weights ~$\Delta$, $3-\Delta$, and
they are related by a  boundary two-point function.\footnote{The conjugated
fields are called 'shadow' fields in the physics literature, cf. \cite{FGGP},
which terminology  was revived in AdS/CFT by Metsaev \cite{Met}.}

The paper is organized as follows.  In Section 2 we give the
preliminaries on the anti de Sitter algebra, its elementary representations,
 the vector-field realizations on the boundary and in the bulk.
In Section 3 we consider the eigenvalue problem in the bulk and we
construct the two-point function in the bulk.
   In Section 4 we give the bulk-to-boundary operator and
construct the integral boundary-to-bulk operator. In Section 5 we
establish the intertwining properties of the boundary-to-bulk and
bulk-to-boundary operators. We display also the intertwining
relation between the two bulk-to-boundary limits of a bulk field.

\setcounter{equation}{0}
\section{Preliminaries}

\subsection{Lie algebra and group}

We need some well-known preliminaries to set up our notation and conventions.
The Lie algebra ~$\cg ~=~  so(3,2)$~ may be defined as the set of $5\times 5$ matrices ~$X$~
which fulfil the relation:\footnote{For other purposes it may be more convenient to use
the other fundamental representation in terms of  $4\times 4$ matrices as in \cite{Dobfour}.}
\beq
  {}^tX\eta + \eta X = 0,
  \label{so32condition}
\eeq
where  the metric $\eta$ is given by
\beq
  \eta = (\eta_{AB}) = \rm{diag}(-1,1,1,1,-1), \quad A, B = 0, 1, \cdots, 4
  \label{so32metric}
\eeq
Then we can choose a basis $ X_{AB} = - X_{BA} $ of $\cg$ satisfying the commutation relations
\beq
  [X_{AB}, X_{CD}] = \eta_{AC} X_{BD} + \eta_{BD} X_{AC} - \eta_{AD} X_{BC} - \eta_{BC} X_{AD}.
  \label{so32com}
\eeq

We list the important subalgebras of $\cg$:
\begin{itemize}
\item $\ck = so(3) \oplus so(2)$, ~generators: ~$X_{AB}$ ~:~ $(A,B) \in  \{1,2,3\}, \{0,4\}$, ~~maximal
compact subalgebra;
\item $\cq$, ~generators: ~$X_{AB}$ ~:~ $A \in  \{1,2,3\}, B\in \{0,4\}$, 
non-compact completion of $\ck$;
\item $\ca = so(1,1)$, ~generator: ~$D \doteq X_{34}\,$, ~dilatations;
\item $\cm = so(2,1)$, ~generators: ~$X_{AB}$ ~:~ $(A,B) \in  \{0,1,2\}$, ~Lorentz subalgebra;
\item $\cn$, ~generators: ~$T_{\mu} = X_{\mu 3} + X_{\mu 4}$, ~$\mu = 0,1,2$, ~translations;
\item $\tcn$,  ~generators: ~$C_{\mu} = X_{\mu 3} - X_{\mu 4}$, ~$\mu = 0,1,2$, ~special conformal transformations.
\item $\ch$, ~generators: ~$D, ~X_{12}$, ~~Cartan subalgebra of ~$\cg$;
\end{itemize}
Thus, we have several decompositions:
\begin{itemize}
\item $\cg = \ck \oplus \cq$, ~Cartan decomposition;
\item $\cg = \ck \oplus \ca \oplus \cn$, ~and ~$\cn \ra\tcn$, ~Iwasawa decomposition;
\item $\cg = \cn \oplus \cm \oplus \ca \oplus \tcn$, ~Bruhat decomposition;
\end{itemize}
The subalgebra ~$\cp = \cm \oplus \ca \oplus \tcn$~ is a maximal parabolic subalgebra of ~$\cg$.

Finally, we introduce the corresponding connected Lie groups:\\
~$G ~=~ SO_0(3,2)$ with Lie algebra ~$\cg=so(3,2)$,
~$K ~=~ SO(3) \times SO(2)$~  is the maximal
compact subgroup of $G$, ~$A ~=~ \exp (\ca) ~=~
SO_0(1,1)$~ is abelian simply connected,
~$N ~=~ \exp (\cn) ~\cong~ \tN ~=~ \exp (\tcn)$,~ are abelian simply
connected subgroups of ~$G$~ preserved by the action of ~$A$.
The group  ~$M ~\cong~ SO_0(2,1)$~ (with Lie algebra $\cm$)
commutes with $A$.
The subgroup ~$P = MAN$~ is a  ~{\it maximal parabolic
subgroup}~ of $G$.
Parabolic subgroups are important because the representations
induced from them generate all admissible irreducible
representations of semisimple groups \cite{Lan,KnZu}.

\subsection{Elementary representations}

We use the approach of \cite{Dob} which we adapt in a condensed form here.
We work with  so-called ~{\it elementary representations} (ERs).
They are induced from representations of ~$P ~=~ MAN\,$, where we use finite-dimensional
representations of spin ~$s\in \ha\bbz_+$~ of $M$, (non-unitary) characters of
$A$ represented by the conformal weight $\D$,
and the factor $N$ is represented trivially.
 The data ~$s,\D$~ is enough to determine a weight
~$\L\in\ch^*$, cf. \cite{Dob}.
Thus, we shall denote the ERs by ~$C^\L$. Sometimes we shall write: ~$\L = [s,\D]$.
The representation spaces are $C^\infty$ functions on $G/P$, or equivalently, on the locally isomorphic
group ~$\tN$~ with appropriate asymptotic conditions (which we do not need explicitly,
cf. e.g.  \cite{DoMo}). We recall that $\tN$ is isomorphic to 3D Minkowski space-time ~$\mathfrak{M}$~
whose elements
will be denoted by ~$x = (x_0,x_1,x_2)$, while the corresponding elements of $\tN$
will be denoted by ~$\tn_x\,$. The Lorentzian inner product in ~$\mathfrak{M}$~
is defined as usual:
\beq
  \lg x,x'\rg  \doteq  x_0 x'_0 - x_1 x'_1 - x_2 x'_2\ ,  
  \label{innerpro}
\eeq
and we use the notation $ \bm{x}^2 = \lg x,x \rg\ . $

The representation action is given as follows:
\eqn{rac} (T^\L (g)\vf) (x) ~=~ y^{-\D}\, D^s (m)\, \vf (x') \eeq
the various factors being defined from the local Bruhat decomposition ~$G \cong_{\rm loc}
\tN AMN$~:
\eqn{vars} g^{-1}\,\tn_x ~=~ \tn_{x'}\, a^{-1} m^{-1} n^{-1} \ , \eeq
where $\,y\in\bbr_+$ parametrizes the elements ~$a\in A$,
~$m\in M$,  ~$D^s (m)$~ denotes the representation action of $M$, ~$n\in N$.

In the above general definition ~$\vf (x)$~ are considered as elements of the
finite-dimensional representation space $V^s$ in which act the operators $D^s (m)$.
Following \cite{DMPPT,Dob} we use scalar functions over an extended space
~$\mathfrak{M} \times ~\mathfrak{M}_0$, where ~$\mathfrak{M}_0$~
is a cone parametrized by the  variable $\z~=~ (\z_0, \z_1, \z_2) $ subject to the condition:\[
  \bm{\z}^2 = \lg \z,\z\rg = \z_0^2 - \z_1^2 - \z_2^2 = 0.  
  \label{conditionforz}
\]
The   internal variable $\z$ will  carry the representation ~$D^s$.

The functions on the extended space will be denoted as ~$\vf(x,\z)$.
On these functions the infinitesimal action of our representations looks as follows:
\bea\label{Boundary}
  & & T_{\mu} = \del{\mu}, \quad \del{\mu} \doteq  \frac{\partial}{\partial x_{\mu}}, \quad \mu = 0, 1, 2
  \nn \\
  & & D =  -\sum_{\mu=0}^2  x_{\mu}  \del{\mu} -\Delta,
  \nn \\
  & & X_{01} = x_0 \del{1} + x_1 \del{0} + \s{01}, \quad
      X_{02} = x_0 \del{2} + x_2 \del{0} + \s{02},
  \nn \\
  & & X_{12} = -x_1 \del{2} + x_2 \del{1} + \s{12},
   \\
  & & C_0 =  2 x_0 D + \bm{x}^2 \del{0} - 2 (x_1 \s{01} + x_2 \s{02}),  
  \nn \\
  & & C_1 = -2 x_1 D + \bm{x}^2 \del{1} + 2 (x_0 \s{01} - x_2 \s{12}),  
  \nn \\
  & & C_2 = -2 x_2 D + \bm{x}^2 \del{2} + 2(x_0 \s{02} + x_1 \s{12}).   
  \nn
\eea
where
\beq
  \s{01} = \z_0 \Del{\z_1} + \z_1 \Del{\z_0}, \qquad
      \s{02} = \z_0 \Del{\z_2} + \z_2 \Del{\z_0}, \qquad
      \s{12} = - \z_1 \Del{\z_2} + \z_2 \Del{\z_1},
  \label{so12}
\eeq
and they satisfy the $\cm = so(1,2) $ commutation relations
\beq
  [\s{01},\s{02}] = -\s{12}, \qquad [\s{02},\s{12}] = \s{01}, \qquad
  [\s{12},\s{01}] = \s{02} \ .
  \label{so12comm}
\eeq

The Casimir of ~$\cg$~ is given by:
\eqn{Casimir:def}
  {\cal C} = \frac{1}{2} X_{AB} X^{AB} = - X_{01}^2 - X_{02}^2 +
  X_{12}^2 - D^2 -3D - C_0 T_0 + C_1 T_1 + C_2 T_2
\eeq
and it is constant on our representation $C^\L$~:
\eqn{cst}
\cc \,\vf ~=~ - (\D (\D-3) + s(s+1) ) \, \vf ~=~ \l(s,\D)\, \vf \ .\eeq
Note that the constant $\l(s,\D)$ has the same value if we replace
 $\D$ by $3-\D$. This  means that the two boundary (shadow) fields
 with conformal weights $\D$ and $3-\D$ are related, or in
mathematical language, that the corresponding representations are
(partially) equivalent.\footnote{We remind that two representations are called ~{\it partially equivalent}~
if there exists an intertwining operator between the two representations. The representations
are called ~{\it equivalent}~ if the intertwining operator is onto and invertible.}

\subsection{Bulk representations}

It is well known that the group ~$SO(3,2)$~ is called also anti de Sitter group, as it is
the group of isometry of 4D anti de Sitter space:
\eqn{adss} \xi^A \,\xi^B\, \eta_{AB} ~=~ 1 \ . \eeq
There are several ways to parametrize anti de Sitter space. We shall utilize the same local Bruhat
decomposition that we used in the previous subsection. Thus, we use the local coordinates on
the factor-space ~$G/MN \cong_{\rm loc} \tN A$, i.e., the coordinates ~$(x,y) ~=~
(x_0,x_1,x_2,y)$, ~$y\in\bbr_+\,$. In this setting anti de Sitter space is called ~{\it bulk}~ space,
while 3D Minkowski space-time is called ~{\it boundary}~ space, as it is identified with the
bulk boundary value ~$y=0$. The functions on the bulk extended with the cone $\mathfrak{M}_0$
will be denoted by ~$\phi (x,y,\z)$.


As we explained in the Introduction we first concentrate on   the holography
principle, or boundary-to-bulk correspondence,  which means to have
an operator which maps a boundary field $\varphi$ to a bulk field
$\phi\,$.
This map must be invariant w.r.t. the Lie algebra $so(3,2)$. In particular, this means
that the Casimir must have the same values in the boundary and  bulk representations.
The Casimir on the boundary representation ~$C^\L$~ is a constant $\l(s,\D)$ given in
\eqref{cst}. Clearly, the (partially) equivalent bulk representation ~$\hC^\L$~ will consist only of functions
on which the Casimir has the same value.

To give more precisely the ~$\hC^\L$~ we first
give a vector-field realization of $ so(3,2) $ on the bulk functions ~$\phi (x,y,\z)$~:
\bea
  & & T_{\mu} = \del{\mu}, \quad \mu = 0, 1, 2
  \nn \\
  & & D  = -\sum_{\mu=0}^2 x_{\mu} \del{\mu} -y \del{y},
  \nn \\
  & & X_{01} = x_0 \del{1} + x_1 \del{0} + \s{01}, \quad
      X_{02} = x_0 \del{2} + x_2 \del{0} + \s{02},
  \nn \\
  & & X_{12} = -x_1 \del{2} + x_2 \del{1} + \s{12},
  \label{Bulk} \\
   & & C_0 =  2 x_0 D + (\bm{x}^2+y^2) \del{0} + 2 (y \s{12} - x_1 \s{01} - x_2 \s{02}),  
  \nn \\
  & & C_1 =  -2 x_1 D + (\bm{x}^2+y^2) \del{1} + 2 ( y \s{02} + x_0 \s{01} - x_2 \s{12}),  
  \nn \\
  & & C_2 =  -2 x_2 D + (\bm{x}^2+y^2) \del{2} + 2( - y \s{01} + x_0 \s{02} + x_1 \s{12}),  
  \nn
\eea
One may verify by straightforward but lengthy computation that (\ref{Bulk}) satisfies (\ref{so32com}).

Note that the realization of $ so(3,2) $ on the boundary given in \eqref{Boundary}
may be obtained from \eqref{Bulk} by replacing ~$y \del{y} \to \Delta$~ and then
taking the limit  $ y \to 0$.

Now we find that the Casimir operator is given in the bulk as follows:
\bea
   {\cal C} ~&=&~  {\cal C}_B + {\cal C}_I
  -2y (\s{12} \del{0} - \s{02} \del{1} + \s{01} \del{2} ),
  \label{Cas-bulk} \\
   && {\cal C}_B ~=~ y^2 (-\del{0}^2 + \del{1}^2 + \del{2}^2) - y^2 \del{y}^2
  + 2 y \del{y}\ , \label{Cas-Bulk} \\
&&   {\cal C}_I  ~=~ (-\s{01}^2 - \s{02}^2 + \s{12}^2) \label{Cas-Internal}
\eea
where $ {\cal C}_I $ is the Casimir operator of $ so(1,2) $ in terms of the internal variables.

Since the Casimir in the bulk is not constant but
a differential operator our representation functions will be found
as the Casimir eigenfunctions in the bulk.
Thus, we consider the eigenvalue problem of the Casimir operator of $ so(3,2):$
\beq
  {\cal C} \phi(x,y,\z) = \lambda(s,\D) \phi(x,y,\z) \ , \quad \phi \in \hC^\L \ .
\eeq
In addition, the elements of ~$\hC^\L$~ must fulfil the appropriate
boundary condition:
\eqn{BouBeh}
   \phi(x,y,\z)|_{y\ra 0} \ \longrightarrow \ y^\D   \phi(x,0,\z) \ , \quad \phi \in \hC^\L \ .
  \eeq
  Later we shall see that the elements of
~$\hC^\L$~ fulfil also the boundary condition with $\D \ra 3-\D$
which is natural having in mind the degeneracy of Casimir values for (partially) equivalent
representations ($\D \lra 3-\D$).

\medskip

Next we mention that the realization (\ref{Bulk}) causes the infinitesimal transformation of the bulk coordinates:
\[
  \begin{array}{ccl}
   T_{\mu} &:& x_{\mu} \to x_{\mu} + a,  \\[3pt]
   D &:& x_{\mu} \to (1-a) x_{\mu}, \qquad y \to (1-a) y, \\[3pt]
   X_{0\mu} &:& x_0 \to x_0 + a x_{\mu}, \qquad x_{\mu} \to x_{\mu} + a x_0, \quad \mu = 1,2 \\[3pt]
   X_{12} &:& x_1 \to x_1 + a x_2, \qquad x_2 \to x_2 - a x_1, \\[3pt]
   C_0 &:& x_0 \to x_0 + a (y^2 - x_0^2 - x_1^2 - x_2^2), \qquad x_{1,2} \to (1-2ax_0) x_{1,2}, \\[3pt]
       & & y \to (1-2ax_0) y, \\[3pt]
   C_1 &:& x_1 \to x_1 + a (y^2 + x_0^2 + x_1^2 - x_2^2), \qquad x_{0,2} \to (1+2ax_1) x_{0,2}, \\[3pt]
       & & y \to (1+ 2ax_1) y, \\[3pt]
   C_2 &:& x_2 \to x_2 + a (y^2 + x_0^2 - x_1^2 + x_2^2), \qquad x_{0,1} \to (1 + 2ax_2) x_{0,1}, \\[3pt]
       & & y \to (1+2ax_2)y.
  \end{array}
\]
It follows that every $ SO(3,2) $ invariant of the two points $ (x_{\mu}, y) $ and $ (x'_{\mu},y') $ is a
function of
\beq
u ~=~   \frac{4yy'}{\bm{(x-x')}^2 + (y+y')^2} \ . 
   \label{SO32inv}
\eeq
We set $ x'_{\mu} = 0, \ y' = 1 $ and obtain a one-point variable\footnote{Sometimes in the literature it is called
colloquially 'one-point invariant'.} which shall be very useful below:
\beq
   \hu ~=~ \frac{4y}{\bm{x}^2 + (y+1)^2} \ . \label{u-def}  
\eeq

\setcounter{equation}{0}
\section{Eigenvalue problem and two-point functions in the bulk}

\subsection{Eigenvalue problem of Casimir in the bulk}

Here we first solve the equation:
\beq
  {\cal C} \Psi(x,y,\z) = \lambda(s,\D) \Psi(x,y,\z) \
\eeq
We are interested in  solutions in which the $\z$-dependence is factored out in the form
\[
  \Psi = \psi(x,y)\, Q(x,y,\z)^s\
\]
where ~$Q$~ is homogeneous  in ~$\z$~ of first degree (which  is due to the fact that  $\Psi$ is homogeneous
in ~$\z$~ of degree ~$s\in\bbz_+$).
We assume that $ \psi $ is a $ SO(3,2)$ invariant, thus it is function  only of $\hu$:
~$\psi(x,y) = \psi(\hu)$.
When $ {\cal C}_B $ acts on $ \psi $ one may write $ {\cal C}_B $ in terms of $\hu$ only:
\beq
 {\cal C}_B = \hu^2 (\hu-1)  \frac{d^2}{d\hu^2} + 2\hu \frac{d}{d\hu} \ .
 \label{CBinu}
\eeq

Furthermore we require that  $ Q $ is an eigenfunction of $ {\cal C}_I. $
This will guarantee that the spin part $ Q^s $ is the eigenfunction
of $ {\cal C}_I $ with the correct spin value.  It follows that $
\Psi $ is also an eigenfunction of $ {\cal C}_I: $ \eqn{casi}
  {\cal C}_I \Psi = \lambda_I \Psi = -s(s+1) \Psi \ .
\eeq
With the fixed vector $(\z'_0, \z'_1, \z'_2) $ in the internal space, we use the following Ansatz for $ Q $
\bea
  Q &=& \frac{2\, I_1  - (\bm{x}^2-(y+1)^2)\, I_2  -
  2(y+1)\, I_3} { \bm{x}^2 + (y+1)^2 }\ ,  
  \label{Qdef} \\
&&  I_1 = \lg x,\z\rg\,\lg x,\z'\rg, \qquad I_2 = \lg\z,\z'\rg, \qquad I_3 = \sum_{\mu=0}^2  x_{\mu}
  (\z \times \z')_{\mu} \ , \nn
\eea
where $ \z \times \z' $ is the standard vector product. (Note that ~$I_1,I_2,I_3$~ are the three possible
scalars which are homogeneous of first degree in both ~$\z$ and $\z'$.)
It is easy to verify that
\beq
  {\cal C}_I I_k = -2 I_k \ . \label{CIonI}
\eeq
Thus, we have
\beq
  {\cal C}_I Q = -2 Q \  \label{CIonQ}
\eeq
and one may verify that $ {\cal C}_I Q^s = -s(s+1) Q^s.$  With this
form of $Q$ the eigenvalue problem is reduced to the second order
differential equation: \beq
  \left( (\hu-1)\hu^2 \frac{d^2}{d\hu^2} + 2\hu \frac{d}{d\hu} - s(s+1) \hu \right) \psi(\hu) =
  (\lambda - \lambda_I) \psi(\hu) = \D(3-\D) \psi(\hu) \ .
  \label{eq:phi}
\eeq

 We sketch the derivation of (\ref{eq:phi}).
First we observe that
\beq
 ({\cal C} - {\cal C}_I) \Psi
 =
 \Psi\, \left[\, \psi^{-1} {\cal C}_B \psi + s Q^{-1} ( {\cal C} - {\cal C}_I ) Q
 + 2s \psi^{-1} Q^{-1} A_1 + s(s-1) Q^{-2} A_2
  \right],
 \label{C-CIPsi}
\eeq
where
\begin{eqnarray*}
  & & A_1 =  -y^2 \bigl( \lg \partial \psi, \partial Q \rg 
  +(\del{y}\psi) (\del{y} Q)  \bigr)
          - y \bigl( (\del{0} \psi) \s{12}Q - (\del{1} \psi) \s{02} Q + (\del{2} \psi) \s{01} Q \bigr),  
  \\
  & & A_2 = -y^2 \bigl( \lg \partial Q, \partial Q \rg 
   + (\del{y}Q)^2 \bigr)
         - 2y \bigl( (\del{0} Q) \s{12}Q - (\del{1} Q) \s{02} Q + (\del{2} Q) \s{01} Q \bigr),  
\end{eqnarray*}
where $ \lg \partial \psi, \partial Q \rg = (\partial_0 \psi) \partial_0 Q - (\partial_1 \psi) \partial_1 Q - (\partial_2 \psi) \partial_2 Q. $ 
Straightforward computation shows that
\[
 ({\cal C}-{\cal C}_I) Q = -2 \hat{u} Q, \qquad  A_1 = 0, \qquad A_2 = -\hu\, Q^2 \ .
\]
Then (\ref{eq:phi}) follows from these and (\ref{CBinu}).

Note that although the derivation is different equation
\eqref{eq:phi} is the same as (7.37) of  \cite{DMPPT}
if we make the change:~ 
 $\D = \ell+2$.


\subsection{Two-point Green function in bulk}

We  need also the two-point Green function in bulk. Standardly for this
 we derive the Green function of the operator ~$ \Cas-\lambda $
\beq
  (\Cas-\lambda) \G(x,y,\z; x',y',\z') =
  y^4 \delta^3(x-x') \delta(y-y') (\z,\z')^s.
  \label{GreenDef}
\eeq
The computation of $ \G $ is more or less same as the one for eigenvalue
problem of $ \Cas $ in the previous subsection. We assume $ \G $ has a factored form
\[
 \G(x,y,\z; x',y',\z') = f(u)\, Q(x,y,\z; x',y',\z')^s,
\]
where $ u$ is the $ SO(3,2)$ invariant of two points $ (x,y) $ and
$ (x',y') $ given in \eqref{SO32inv}. This assumption is justified a posteriori
since the two-point function is unique up to multiplicative constant.

   Note that ~$\G$~  is a eigenfunction of $ \Cas_I, $ i.e.,
$ \Cas_I \G = \lambda_I \G. $ Then $ \G $ is given by
\bea
  & & \G = u^{\D}\, F(u)\, Q^s \ ,
   \nn \\
  & & Q = \frac{ 2\,I'_1 - ((\bm{x}-\bm{x}')^2-(y+y')^2)\, I_2 - 2(y+y')\, I'_3}
               {  (\bm{x}-\bm{x}')^2 + (y+y')^2 }\ , \\ 
  \label{GreenSol}
 &&  I'_1 = \lg x-x',\z\rg\, \lg x-x',\z'\rg, \quad I_2 = \lg\z,\z'\rg, \quad
  I'_3 = \sum_{\mu=0}^2 (x_{\mu}-x'_{\mu}) (\z \times \z')_{\mu},\nn
\eea
and $ F(u) $ is a singular solution of the hypergeometric equation
\beq
 \left( u(1-u) \frac{d^2}{du^2} + 2[\D-1 - \D\, u] \frac{d}{du} + (s -\D +1)(s+\D) \right) F(u) = 0.
 \label{GreenF}
\eeq

 We sketch the derivation of the Green function.
First we observe that
\beq
 ({\cal C} - {\cal C}_I) \G
 =
 \G\, \left[\, f^{-1} {\cal C}_B f + s Q^{-1} ( {\cal C} - {\cal C}_I ) Q
 + 2s f^{-1}  Q^{-1} A_1 + s(s-1) Q^{-2} A_2
  \right],
 \label{C-CIPsi2S}
\eeq
where
\begin{eqnarray*}
  & & A_1 =  -y^2 \bigl( \lg \partial f, \partial Q \rg +(\del{y}f) (\del{y} Q)  \bigr)
          - y \bigl( (\del{0} f) \s{12}Q - (\del{1} f) \s{02} Q + (\del{2} f) \s{01} Q \bigr), 
  \\
  & & A_2 = -y^2 \bigl( \lg \partial Q, \partial Q \rg + (\del{y}Q)^2 \bigr)
         - 2y \bigl( (\del{0} Q) \s{12}Q - (\del{1} Q) \s{02} Q + (\del{2} Q) \s{01} Q \bigr). 
\end{eqnarray*}
Straightforward computation shows that
\[
  (\Cas-\Cas_I) Q = -2u Q, \qquad
  A_1 = 0, \qquad A_2 = -u Q^2.
\]
It follows that
\[
  (\Cas-\Cas_I) \G = Q^s \left(
     u^2(u-1) \frac{d^2}{du^2} + 2u \frac{d}{du} - s(s+1) u \right) f(u).
\]
Setting $ f(u) = u^{\D} F(u) $ we have
\[
  (\Cas - \lambda) \G = -Q^s  u^{\D+1}
  \left(
     u(1-u) \frac{d^2}{du^2} + 2[\D-1 - \D\, u] \frac{d}{du} + (s -\D +1)(s+\D)
  \right) F(u).
\]
Thus if $ F(u) $ is a singular solution of the hypergeometric equation
then we obtain the RHS of (\ref{GreenDef}). The delta functions in the RHS
corresponds to the singularity at $ u = 1 \Leftrightarrow x_{\mu}=x_{\mu}', \ y = y'. $

\medskip

\noindent{\bf Remark 1:} One may wonder whether the above may be generalised to
the anti de Sitter algebra $so(d,2)$ for  $d>3$. Actually, the only difficulty for $d>3$ and nontrivial spin would be
to find the  explicit form of the function ~$Q$.  In fact, below some calculations are valid implicitly
or explicitly for arbitrary $d$.


\setcounter{equation}{0}
\section{Bulk-boundary correspondence}

Consider the fields ~$\varphi \in C^\L$~ and ~$\phi\in \hC^\L$~ from  the boundary and
 bulk representations for the same $\L$. By construction  they are eigenfunctions of the Casimir operator
with the same eigenvalue:
\beq
  \Cas \varphi = \lambda \varphi, \qquad \Cas \phi = \lambda \phi.
  \label{fields}
\eeq
The bulk field behaves as in \eqref{BouBeh}  when approaching  the boundary.
Thus, we define the bulk-to-boundary operator $ L_\D $ by:
\eqnn{Lop}
&& L_\D ~:~ \hC^\L ~\longrightarrow ~C^\L \\
&& \varphi(x,\z) = (L_\D \phi)(x,\z) := \lim_{y \to 0} y^{-\D}
  \phi(x,y,\z). \nn
\eea

 On the other hand the boundary-to-bulk operator $\tilde{L}_\L$ is defined by:
\eqnn{Ltildeop}
&& \tilde{L}_\L  ~:~ C^\L ~\longrightarrow ~\hC^\L \\
&&  \phi(x,y,\z) = (\tilde{L}_\L \varphi)(x,\z) :=
  \int S_\L(x-x',y ; \z,\partial_{\z'}) \varphi(x',\z') d^3x'
  \nn
\eea
where the kernel $ S_\L $ is obtained from the two-point Green function
$ \G $ defined in (\ref{GreenDef}) as follows
\beq
   S_\L(x-x',y ; \z,\partial_{\z'}) ~=~
   \lim_{y' \to 0} y'{}^{\D-3} \G(x,y,\z;x',y',\partial_{\z'}).
   \label{G2S}
\eeq
The formula for $ S_\L $ (for $s\in\bbz_+$) is given by
\beq
  S_\L ~=~ N_\L\, \tu^{3-\D} R^{s}, \quad \tu = \frac{4y}{(\bm{x}-\bm{x}')^2 + y^2}, \quad
  R = \frac{\tu\cl}{4y},  
  \label{S-formula}
\eeq
with
\bea
  & & \cl = 2 \tI_1 - ( (\bm{x}-\bm{x}')^2 - y^2 ) \tI_2 - 2y \tI_3,
  \label{Lambda-def} \\
  & & \tI_1 = \lg x-x',\z\rg\, \lg x-x',\del{\z'}\rg, \quad \tI_2 = \lg\z, \del{\z'}\rg, \quad
      \tI_3 = \sum_{\mu=0}^2 (x_{\mu}-x'_{\mu}) (\z \times \del{\z'})_{\mu}\ ,
  \nn\eea
and ~$N_\L$~ is a normalization constant depending on the representation ~$\L = [s,\D]$.

\medskip

Now we check consistency of the operators $ L_\D $ and $ \tilde{L}_\L$:
\eqn{consi} L_\D \circ \tilde{L}_\L ~=~ \mbox{\bf 1}_{\mbox{$C^\L$}} \ ,
\qquad
 \tilde{L}_\L \circ L_\D ~=~ \mbox{\bf 1}_{\mbox{$\hC^\L$}} \ . \eeq

For the first relation in \eqref{consi} we have to show that:
\bea
  \varphi(x,\z) &=& (L_\D \circ \tilde{L}_\L \varphi)(x,\z) \nn \\
  &=& \lim_{y \to 0} y^{-\D} \int S_\L(x-x',y ; \z,\partial_{\z'}) \varphi(x',\z') d^3x'.
  \label{consis}
\eea
We take the limit first by exchanging it and the integral.
To calculate the limit it is necessary to express the kernel $ S_\L $ in another form.
To this end we establish the following formula of Fourier transform:
\beq
  \int \frac{e^{i\lg p,X\rg}}{(\lg X,X\rg + y^2)^{\alpha}} \frac{d^3X}{(2\pi)^{3/2}}
  =
  \frac{ i\pi }{ (-1)^{2\alpha-1} 2^{\alpha} \Gamma(\alpha) } \left( \frac{ \sqrt{-\bm{p^2}}\, }{ y } \right)^{\alpha-3/2}  H_{\alpha-3/2}^{(1)}
  (y\sqrt{-\bm{p^2}}\,), 
  \label{Ftrans}
\eeq
where $ X_{\mu} = x_{\mu}-x'_{\mu} $ and $ H_{\b}^{(1)} $ is a Hankel function.
The $ (X_1,X_2)$ integration can by carried out by making use of the following two formulae:
First one is a formula for $(d-1)$ dimensional angular integration in $d$ dimensional Euclidean space:
\bea
 & &
 \int f(r) e^{-i\vec{p}\cdot \vec{x} } \frac{d^dx}{(2\pi)^d} =
 \left( \frac{1}{2\pi} \right)^{d/2} \int_0^{\infty} r f(r)
 \left( \frac{r}{p} \right)^{\frac{d}{2}-1} J_{\frac{d}{2}-1}(pr)\, dr\ ,
 \label{ang-int} \\
 & & \vec{p} = (p_1, p_2, \cdots, p_d), \ \vec{x} = (x_1, x_2, \cdots, x_d), \ p^2 = \sum_{k=1}^d p_k^2, \
     r^2 = \sum_{k=1}^d x_i^2, \nn
\eea
which is valid for any radial function $f(r). $
Second formula is an integration of Bessel function:
\beq
  \int_0^{\infty} \frac{ r^{\b+1} J_{\b}(ar) }{ (r^2 + \rho^2)^{\g+1} } dr =
  \frac{ a^{\g} \rho^{\b-\g} K_{\b-\g}(a\rho) }{ 2^{\g} \Gamma(\g+1) }\ , \quad
  2\Re\, \g + \frac{3}{2} > \Re\, \b > -1.
  \label{int-Bessel}
\eeq
We modify the second formula (\ref{int-Bessel}).
Set $ \b = 0 $ and replace $ \rho $ with $ -i\rho, $ then use the relation
between Bessel functions
\beq
   K_{\g}(z) = \frac{\pi}{2} i e^{\g\pi i/2} H_{\g}^{(1)}(iz), \quad
   -\pi < \arg z < \frac{\pi}{2}
   \label{Bessel-Bessel}
\eeq
we obtain
\beq
  \int_0^{\infty} \frac{ r J_0(ar) }{ (r^2-\rho^2)^{\g+1} } dr
  =
  \frac{  i \pi a^{\g} }{(-1)^{\g} 2^{\g+1} \Gamma(\g+1) \rho^{\g} } H_{\g}^{(1)}(a\rho).
  \label{int-Bessel2}
\eeq
Now we return to the Fourier transform (\ref{Ftrans}). Angular integration in
$X_1X_2$ plane is performed by (\ref{ang-int}) and we use
(\ref{int-Bessel2}) for the radial integration in the plane:
\begin{eqnarray*}
 & &
  \int \frac{e^{i\lg p,X\rg}}{(\lg X,X\rg + y^2)^{\alpha}} \frac{d^3X}{(2\pi)^{3/2}}
  = \frac{1}{ \sqrt{2\pi} } \int_{-\infty}^{\infty} dX_0 \int_0^{\infty} dr
  \frac{ r J_0(\tilde{p}r) }{ (-1)^{\alpha} (r^2-X_0^2-y^2)^{\alpha} } e^{ip_0X_0}
 \\[3pt]  
 & &
  = \frac{i\pi}{ (-1)^{2\alpha-1} \sqrt{2\pi} \Gamma(\alpha) }
    \left( \frac{\tilde{p}}{2} \right)^{\alpha-1}
    \int_0^{\infty} dX_0 \frac{ H_{\alpha-1}^{(1)}(\tilde{p}\sqrt{X_0^2+y^2}\,) }
    { (X_0^2+y^2)^{(\alpha-1)/2} }
    \cos p_0 X_0
\end{eqnarray*}
where
\[
  r^2 = X_1^2+X_2^2, \qquad \tilde{p}^2 = p_1^2 + p_2^2.
\]
Recalling that
\[
  H^{(1)}_{\b}(z) = J_{\b}(z) + i Y_{\b}(z)
\]
$X_0$ integration is performed by the formulae of Fourier cosine transform \cite{Bateman}
\[
 \begin{array}{l|l}
   f(r) & \int_0^{\infty} f(r) \cos (r\rho)\, dr \\ \hline
   \displaystyle
   \frac{ J_{\b}(a \sqrt{r^2+b^2}\,) }{ (r^2+b^2)^{\b/2} }
   & \displaystyle \left\{
     \begin{array}{ll}
     \displaystyle
     \sqrt{ \frac{\pi b}{2} } \frac{ (a^2-\rho^2)^{\b/2-1/4} }{ (ab)^{\b} } J_{\b-1/2}(b\sqrt{a^2-\rho^2}\,)
     & 0 < \rho < a \\[15pt]
     0 & a < \rho
     \end{array}
     \right.
   \\ \hline
    \displaystyle
     \frac{ Y_{\b}(a \sqrt{r^2+b^2}\,) }{ (r^2+b^2)^{\b/2} }
    & \displaystyle \left\{
     \begin{array}{ll}
     \displaystyle
     \sqrt{ \frac{\pi b}{2} } \frac{ (a^2-\rho^2)^{\b/2-1/4} }{ (ab)^{\b} } Y_{\b-1/2}(b\sqrt{a^2-\rho^2}\,)
     & 0 < \rho < a \\[15pt]
      \displaystyle
      -\sqrt{ \frac{2 b}{\pi} } \frac{ (\rho^2-a^2)^{\b/2-1/4} }{ (ab)^{\b} } K_{\b-1/2}(b\sqrt{\rho^2-a^2}\,)
     & a < \rho
     \end{array}
     \right.
 \end{array}
\]
\hfill
$
 \Re\, \b > -\frac{1}{2}, \quad a,  b > 0
$

\noindent
By these formula we obtain
\begin{eqnarray}
 & &
    \int_0^{\infty} dX_0 \frac{ H_{\alpha-1}^{(1)}(\tilde{p}\sqrt{X_0^2+y^2}\,) }
    { (X_0^2+y^2)^{(\alpha-1)/2} }    \cos (p_0 X_0)
 \nn \\[3pt]
 & &
  \qquad =
  \left\{
    \begin{array}{lcl}
      \displaystyle
        \left( \frac{\pi y}{2} \right)^{1/2} \frac{ (- \lg p,p\rg)^{\alpha/2-3/4} }{(\tilde{p}y)^{\alpha-1} } H^{(1)}_{\alpha-3/2}
        (y\sqrt{\tilde{p}^2-p_0^2}\,)
        & & 0 < p_0 < \tilde{p}
      \\[15pt]  
      \displaystyle
        -i \left( \frac{2 y}{\pi} \right)^{1/2} \frac{\lg p,p\rg^{\alpha/2-3/4} }{(\tilde{p}y)^{\alpha-1} } K_{\alpha-3/2}
        (y\sqrt{p_0^2-\tilde{p}^2}\,)
        & &   \tilde{p} < p_0  
    \end{array}
  \right.
  \nn \\[3pt]
 & &
  \qquad =
   \left( \frac{\pi y}{2} \right)^{1/2} \frac{ (-\lg p,p\rg)^{\alpha/2-3/4} }{(\tilde{p}y)^{\alpha-1} } H^{(1)}_{\alpha-3/2}
   (y\sqrt{\tilde{p}^2-p_0^2}\,).
   \label{X0-integral}  
\end{eqnarray}
In the last equality the relation (\ref{Bessel-Bessel}) was used to unify two separate cases.
Note that $ \lg p,p\rg =  p_0^2- \tilde{p}^2.$ 
In this way the Fourier transform (\ref{Ftrans}) has been established.

 Now we evaluate the Fourier transform of the kernel $ S_\L$
\bea
 & &
  \int  S_\L(X,y ; \z,\partial_{\z'}) e^{i\lg p,X\rg} \frac{d^3X}{(2\pi)^{3/2}} 
  =
 N_\L \int \frac{(4y)^{3-\D}}{ (\lg X,X \rg + y^2)^{s-\D+3} }\ \bm{S}^{s}\ e^{i\lg p,X\rg} \frac{d^3X}{(2\pi)^{3/2}}
 \nn \\[3pt] 
 & &
 \qquad =
 \frac{ - i\pi N_\L}{2^{s+\D-1}  \Gamma(s-\D+3) y^{s-3/2}} \bm{S}^{s}
 ( \sqrt{-\bm{p^2}}\, )^{s-\D+3/2}  H_{s-\D+3/2}^{(1)}(y\sqrt{-\bm{p^2}}\,),
 \nn   
\eea
where
\[
 \bm{S} = -2 (\del{p} \cdot \z) \, \del{p}\cdot\del{\z'}  +
 (\lg\del{p},\del{p}\rg+y^2)\lg\z,\del{\z'}\rg
 +2iy \lg\del{p}, \z\times \del{\z'}\rg,  
\]
with $ \displaystyle a\cdot b = \sum_{\mu=0}^2 a^{\mu} b_{\mu}. $
Inverse Fourier transform gives the following formula of the kernel
\beq
 S_\L =
 \frac{ -i\pi N_\L }{2^{s+\D-1}  \Gamma(s-\D+3) y^{s-3/2}} \int \bm{S}^{s}
 ( \sqrt{-\bm{p^2}}\, )^{s-\D+3/2}  H_{s-\D+3/2}^{(1)}(y\sqrt{-\bm{p^2}}\,) e^{-i(\lg p,X\rg}
  \frac{d^3p}{(2\pi)^{3/2}}.
  \label{S-newform}   
\eeq
Since we take a limit of $ y \to 0, $ we replace the Hankel function with its asymptotic form
\[
  -i H_{\alpha}^{(1)}(z) \to -\frac{\Gamma(\alpha)}{\pi}  \left( \frac{2}{z} \right)^{\alpha},
  \qquad z \to 0
\]
Then
\[
 -i ( \sqrt{-\bm{p^2}}\, )^{s-\D+3/2}  H_{s-\D+3/2}^{(1)}(y\sqrt{-\bm{p^2}}\,) 
 = -\frac{\Gamma(s-\D+3/2)}{\pi}
 \left( \frac{2}{y} \right)^{s-\D+3/2} \ , \quad s-\D+3/2 \notin \bbz_- \ ,
\] is independent of $p_{\mu} $ so that the action of $ \bm{S} $ is reduced to
$ y^2 \lg\z,\del{\z'}\rg  $ and  
the integration over $p$ becomes Dirac's delta function:
\beq
  S_\L ~\ra~  -\frac{ (2\pi)^{3/2}\,N_\L\, \Ga(s-\D+3/2) }{ 2^{2\D -5/2}\, \Ga(s-\D+3)}  
  y^{\D} \delta^3(X) \lg \z,\del{\z'}\rg^{s}\ , \qquad s-\D+3/2 \notin \bbz_-  \ , ~~~y \to 0 \ .
  \label{S-newform2}
\eeq
Substituting this formula of $ S$ in \eqref{consis} we obtain:
\bea\label{consist}  \varphi(x,\z) ~&=&~
-\frac{ \pi^{3/2}\, N_\L \, \Ga(s-\D+3/2) }{ 2^{2\D -4}\, \Ga(s-\D+3)}  
    \  \varphi(x,\z) , \\
&&  s-\D+3/2 \notin \bbz_- \ , \quad s-\D+3 \notin \bbz_- \ .
\nn\eea
From the latter we see the first consistency relation (\ref{consi})
being true by an appropriate choice of ~$N_\L\,$, e.g.
\bea\label{norma} N_\L ~&=&~ -\frac{  2^{2\D -4}\, \Ga(s-\D+3)}{ \pi^{3/2}\,  
 \Ga(s-\D+3/2) } \ ,\\
 &&  s-\D+3/2 \notin \bbz_- \ , \quad s-\D+3 \notin \bbz_- \ .
\nn\eea

As a ~{\it Corollary}~ we conclude that for generic values of ~$\D$~
we can reconstruct a function on anti de Sitter space from its boundary
value. Indeed, suppose we have:
\eqn{rst} \phi(x,y,\z) ~=~ \int S_\L (x-x',y;\z,\del{\z'}) f(x',\z') \,d x' \ . \eeq
Then we have for the boundary value:
\bea\label{rss}  \psi_0(x,\z) ~&\doteq&~ (L_\D\, \phi) (x,\z) ~=~
\lim_{y\to 0}\ y^{-\D}\ \phi(x,y,\z)
~=\nn\\ =&&~ \lim_{y\to 0}\ y^{-\D}\
\int S_\L (x-x',y;\z,\del{\z'}) f(x',\z') \,d x'
~=~ f(x) \eea

\medskip

Now we can prove the second consistency relation in \eqref{consi}:
\bea\label{invd}   \big( {\tilde L}_\L \circ L_\D\, \phi \big)
(x,y,\z) ~&=&~
\int S_\L (x-x',y;\z,\del{\z'}) \big(  L_\D\, \phi \big) (x',\z') \,d x' ~=\cr
~&=&~ \int S_\L (x-x',y;\z,\del{\z'})
\lim_{y'\to 0}\ y'^{{-\D}}\,  \phi (x',y',\z') ~=\cr
&=&~  \int S_\L (x-x',y;\z,\del{\z'})  \, \psi_0 (x',\z') \,d x' ~=~ \phi (x,y,\z) \eea
where in the last line we used \eqref{rss}.

\setcounter{equation}{0}
\section{Intertwining properties}

Here we investigate the intertwining properties of the boundary $ \leftrightarrow  $ bulk operators.

\subsection{Bulk-to-boundary operator ~$L_\D$}

It is not difficult to verify the intertwining property of the Bulk-to-boundary operator ~$L_\D$.
  Namely, one should verify the following:
\beq
  L_\D \circ \hat{X} = \tilde{X} \circ L_\D\ , 
  \label{IT1}
\eeq
where ~$X \in so(3,2)$, ~$ \tilde{X} $ denotes the action of the generator $X$ on the boundary (\ref{Boundary})
and $ \hat{X} $ denotes the action of the generator in the bulk (\ref{Bulk}).
More explicitly,
\beq
 \tilde{X}\, \varphi(x,\z) = \lim_{y \to 0} y^{-\D}\, \hat{X}\, \phi(x,y,\z)\ , \quad \vf \in C^\L \ , ~
 \phi \in \hC^\L
 \label{IT1_2}
\eeq

If the field $\vf$ belongs to the conjugate representation ~$\vf\in C^{\tilde\L}$,
~$\tL = [s,3-\D]$, then relations \eqref{IT1},\eqref{IT1_2} hold with the change
~$\D \to 3-\D$, the same change being made also in (\ref{Boundary}).

  \subsection{Boundary-to-bulk operator  $\tilde{L}_\L$}

 The intertwining property of the boundary-to-bulk operator  ~$\tilde{L}_\L$~ means that
\beq
   \hat{X} \circ \tilde{L}_\L = \tilde{L}_\L\circ \tilde{X} \
   \label{IT2}
\eeq
More explicitly, it reads
\beq
  \hat{X}\, \phi(x,y,\z) =
  \int S_\L(x,y,\z;x',\partial_{\z'})
  \,\tilde{X}_\L\, \varphi(x',\z') d^3x', \quad \phi \in \hC^\L , ~\vf \in C^\L \
  \label{IT2_2}
\eeq

This is an immediate consequence of  ~$L_\D \circ \tilde{L}_\L = \mbox{\bf 1}_{\mbox{$C^\L$}}$,
~$ \tilde{L}_\L \circ L_\D = \mbox{\bf 1}_{\mbox{$\hC^\L$}} $~
and (\ref{IT1}). By sandwiching (\ref{IT1}) by $ \tilde{L}_\L $ one has
\[
  \tilde{L}_\L \circ L_\D \circ \hat{X} \circ \tilde{L}_\L =
  \tilde{L}_\L \circ \tilde{X} \circ L_\D \circ \tilde{L}_\L \ , \quad   {\rm acting~on} ~C^\L
\]
This is nothing but (\ref{IT2}).

\bigskip\noindent
{\bf Proof of (\ref{IT2_2}) by direct computation}

A key observation to check the intertwining property is the following identities:
\beq
 \frac{\partial \tu}{\partial x_{\mu}} = -\frac{\partial \tu}{\partial x'_{\mu}},  \qquad
 \frac{\partial R}{\partial x_{\mu}} = -\frac{\partial R}{\partial x'_{\mu}},  \quad \mu =0, 1, 2  \label{identity1}
\eeq
It follows that
\beq
 \frac{\partial S_{\Lambda}}{\partial x_{\mu}} = -\frac{\partial S_{\Lambda}}{\partial x'_{\mu}},  \quad \mu =0, 1, 2
 \label{identity2}
\eeq
Formulas of differentiation by $y:$
\[
  \frac{\partial \tu}{\partial y} = \frac{\tu}{y}-\frac{\tu^2}{2}, \qquad
  \frac{\partial R}{\partial y} = \frac{\tu}{4y}\del{y}{\cal L} - \frac{\tu R}{2}.
\]
It follows that
\bea
 & & \frac{\partial S_{\Lambda}}{\partial y} =
 \left( \frac{3-\D}{y} - (s-\D+3) \frac{\tu}{2}+
 s\frac{1}{\cal L} \del{y}{\cal L} \right) S_{\Lambda},
  \nn \\
  & &
  \left( \sum_{\mu=0}^2 x_{\mu} \frac{\partial}{\partial x_{\mu}} + y \frac{\partial}{\partial y} \right) S_{\Lambda}
  = - \left(3 - \D +\sum_{\mu=0}^2 x'_{\mu} \frac{\partial}{\partial x'_{\mu}} \right) S_{\Lambda}.
  \label{Sbyy}
\eea

With this identity, it is immediate to verify the intertwining property for $ T_{\mu} $ and $ D $.

\subsection{Further intertwining relations}

 We start by recording the second limit of the bulk functions
\bea\label{sndl}  \varphi_0(x,\z) ~&\doteq&~
\lim_{y\to 0}\ y^{\D-3}\    \phi(x,y,\z)
~=\\ &=&~ \lim_{y\to 0}\ y^{\D-3}\
\int S_\L (x-x',y;\z,\del{\z'})\, \psi_0(x',\z') \,d x'
~=\nn\\   &=&~
N_\L\ \lim_{y\to 0}\ y^{\D-3}\ \int \left( \frac{4y}{(\bm{x}-\bm{x}')^2 + y^2} \right)^{3-\D}
\left(\frac{\cl}{{(\bm{x}-\bm{x}')^2 + y^2}} \right)^s\, \psi_0(x',\z') \,d x'  
 \nn\\ &&
\nn\\   &=&~
N_\L\ \int \left( \frac{4}{(\bm{x}-\bm{x}')^2} \right)^{3-\D}
\left(\frac{ 2 \tI_1 - ( (\bm{x}-\bm{x}')^2  ) \tI_2}{{(\bm{x}-\bm{x}')^2 }} \right)^s  
\, \, \psi_0(x',\z') \,d x' ~=\nn\\
&=&~
\cn_\L\ \int {d x'\over {((\bm{x}-\bm{x}')^2)^{3-\D}}}\  
\left( \frac{2 \lg x-x',\z\rg\, \lg x-x',\del{\z'}\rg}{{(\bm{x}-\bm{x}')^2 }}
- \lg\z, \del{\z'}\rg \right)^s\ \psi_0(x',\z')   
\nn\\
&=&~
\cn_\L\ \int {d x'\over {((\bm{x}-\bm{x}')^2)^{3-\D}}}\  
 \left(r(x-x';\z,\del{\z'})\right)^s\  \psi_0 (x',\z')
~=\nn\\
  &=&~ \frac{\cn_\L}{\g_{\tL}}
\   \int d x'\ G_{\tL} (x-x';\z,\del{\z'})\
\psi_0 (x',\z') \ , \qquad \cn_\L = 4^{3-\Delta} N_\L, \nn  
\eea
where in the second line we have used \eqref{rss}, in the third line
we have used \eqref{S-formula} and \eqref{Lambda-def}, and in the last
line we have recovered the well-known conformal two-point function, cf., e.g.,
\cite{Pol}:
\bea\label{twop} G_\L (x;\z,\z') ~&=&~ \g_\L\ \frac{(r(x;\z,\z'))^s}
{(\bm{x}^2)^{\D}}\ , \\   
 r(x;\z,\z') ~&=&~ r(x)_{\mu\sigma}\z^\mu\z'^\sigma \ , \quad
 r(x)_{\mu\sigma}~=~ \frac{2}{\bm{x}^2} x_\mu x_\sigma - g_{\mu\sigma} \nn \\ 
  g ~&=&~ (g_{\mu\nu}) = \rm{diag}(1,-1,-1) \nn 
\eea
for the conjugate weight ~$\tL = [s,3-\D]$. The latter is natural since
~$\psi_0 \in C^\L$, ~$\varphi_0 \in C^{\tL}$,  and the conformal two-point function
realizes the equivalence of the conjugate representations ~$\L,\,\tL$~
which have the same Casimir values, cf. \cite{DMPPT}.
The normalization constant ~$\g_\L$~ depends on the representation ~$\L = [s,\D]$~
and below we derive a formula for the product ~$\g_\L\,\g_\tL\,$.

Further, using \eqref{sndl}  we define the operator ~$G_\L$~ through the kernel ~$G_\L (x;\z,\z')$~:
\eqnn{twopo} G_\L ~&:&~ C^\tL ~\ra~ C^\L \ , \\
(G_\L \varphi_0) (x,\z) ~&=&~ \int d x'\ G_{\L} (x-x';\z,\del{\z'})\
\varphi_0 (x',\z') \ . \nn\eea
Then relation \eqref{sndl} may be written as:
\eqn{sndla} L_{\tD} ~=~ \frac{\cn_\L}{\g_{\tL}}\ G_\tL \circ L_\D\ , 
\qquad \tD \doteq 3-\D \ \eeq
as operators acting on the bulk representation ~$\hC^\L$. 

Note that at generic points (those not excluded in \eqref{consist}) the
operators ~$G_\L$~ and ~$G_\tL$~ are inverse to each other \cite{DMPPT}:
\eqn{inveg} G_\L \circ G_\tL ~=~ \mbox{\bf 1}_{C^\L} \ ,
\qquad G_\tL \circ G_\L ~=~ \mbox{\bf 1}_{C^\tL} \ . \eeq

At generic points from this we can obtain a lot of
interesting relations, e.g., applying ~$\tilde L_\L$~
from the right we get:
\eqn{sss}  L_\tD \circ \tilde L_\L  ~=~
{\cn_\L \over \g_\tL}\ G_\tL \eeq  
Then we write down the conjugate relation:
\eqn{sssz}  L_\D \circ \tilde L_\tL  ~=~
{\cn_\tL \over \g_\L}\ G_\L \eeq  
Then we combine relations   \eqref{sss} and \eqref{sssz}:
\eqn{zzz} L_\D \circ \tilde L_\tL \circ L_\tD \circ \tilde L_\L  ~=~
{\cn_\tL \over \g_\L}\ {\cn_\L \over \g_\tL}\  G_\L \circ G_\tL
~=~ {\cn_\tL \over \g_\L}\ {\cn_\L \over \g_\tL}\ \mbox{\bf 1}_\L  
\eeq
For the LHS of \eqref{zzz} we use first the second relation of
\eqref{consi}, then the first to obtain:
\eqn{yyy} L_\D \circ \tilde L_\tL \circ L_\tD \circ \tilde L_\L
~=~ L_\D \circ \mbox{\bf 1}_{\hC^\L} \circ \tilde L_\L
~=~ L_\D \circ  \tilde L_\L ~=~ \mbox{\bf 1}_\L \ .
\eeq
Thus, from \eqref{zzz} and \eqref{yyy} follows:
\eqnn{normas} \g_\L\ \g_\tL ~&=&~ \cn_\L\ \cn_\tL ~=~
\frac{ 2^4 \Ga(s-\D+3)\, \Ga(s+\D)}{ \pi^3\,     
 \Ga(s-\D+3/2)\, \Ga(s+\D-3/2)} \ , \\
 &&  s-\D+3/2 \notin \bbz_- \ , \quad s-\D+3 \notin \bbz_- \   ,\nn\\
&& s+\D-3/2 \notin \bbz_- \ , \quad s+\D \notin \bbz_- \ .
\nn\eea
The product of constants in \eqref{normas} should be proportional to the
the analytic continuation of the Plancherel measure for the Plancherel formula
contribution of the principal series of unitary irreps of ~$G$, cf., e.g., \cite{Dobads},
but we shall not go into that.

\medskip

\noindent{\bf Remark 2:} One may wonder what happens at the excluded values in \eqref{normas}. This requires
further nontrivial examination. Such study was done in the Euclidean case in \cite{DMPPT}. Since some
results may follow by Wick rotation we may conjecture that, for example,
the operators ~$G_\L$~ and ~$G_\tL$~ would not be inverse to each other. This would be since at these points
the representations ~$C^\L$~ and ~$C^\tL$~ would be reducible and the $G$-operators would have kernels.
All such properties are currently under study \cite{AiDo2}.

\np

\end{document}